\documentclass[letterpaper]{article} 
\usepackage{aaai25}  
\usepackage{times}  
\usepackage{helvet}  
\usepackage{courier}  
\usepackage[hyphens]{url}  
\usepackage{graphicx} 
\usepackage{natbib}  
\usepackage{caption} 
\usepackage{algorithm}
\usepackage{algorithmic}

\usepackage{newfloat}
\usepackage{listings}
\DeclareCaptionStyle{ruled}{labelfont=normalfont,labelsep=colon,strut=off} 
\floatstyle{ruled}
\newfloat{listing}{tb}{lst}{}
\floatname{listing}{Listing}
\title{What Can Youth Learn About Artificial Intelligence and Machine Learning in One Hour? Examining How Hour of Code Activities Address the Five Big Ideas of AI}
\author {
    Luis Morales-Navarro,
    Yasmin B. Kafai,
    Eric Yang,
    Asep Suryana
}
\affiliations {
    University of Pennsylvania\\
    luismn@upenn.edu, kafai@upenn.edu, eriyang@sas.upenn.edu, asepsryn@upenn.edu
}

\usepackage{bibentry}

\begin{document}

\maketitle

\begin{abstract}
The prominence of artificial intelligence and machine learning in everyday life has led to efforts to foster AI literacy for all K–12 students. In this paper, we review how Hour of Code activities engage with the five big ideas of AI, in particular with machine learning and societal impact. We found that a large majority of activities focus on perception and machine learning, with little attention paid to representation and other topics. A surprising finding was the increased attention paid to critical aspects of computing. However, we also observed a limited engagement with hands-on activities. In the discussion, we address how future introductory activities could be designed to offer a broader array of topics, including the development of tools to introduce novices to artificial intelligence and machine learning and the design of more unplugged and collaborative activities.
\end{abstract}

%

\section{Introduction}

The use of artificial intelligence (AI) and machine learning (ML) applications is pervasive in everyday life. Most youth interact with these applications, knowingly or not, when using social media, playing games, or going to school. While most adults and youth see benefits of using these applications \cite{hornberger2023whatdo, skjuve2023theuser}, they have little understanding of how they work, often underestimating or misattributing outcomes \cite{long2020ai, solyst2024children}. To address these shortcomings, there have been increasing calls for an AI literacy for all. However, what it means to be AI literate is an open question, with multiple proposals under discussion, such as understanding the five big ideas \cite{touretzky2019envisioning}, developing various competencies, \cite{long2020ai} or introducing new forms of computational thinking \cite{tedre2021ct}.

Most K–12 computing education efforts do not address AI/ML ideas and competencies in their courses and frameworks. While numerous AI curricula have been developed to introduce K–12 students to these ideas \cite{morales2024unpacking}, implementing these at a wider scale will require extensive teacher preparation and development in a field that is only slowly making its way into K–12 education. As a possible solution, various short introductory activities have been developed, offering learners and teachers free access to tools and learning resources. One of the largest and most popular resources is the Hour of Code, started by the US non-profit Code.org. Other more recent efforts include AI4ALL \cite{judd2020activities} and Day of AI \cite{breazeal2023day} which provide an introduction to key concepts, challenges, and careers in AI/ML.

In this paper, we focus on introductory activities in the Hour of Code (hereafter: HoC) started over a decade ago with the goal of fostering the introduction of computer science in K–12 education. HoC offers a range of introductory, grade-specific computing activities on different platforms, some even unplugged (i.e., using no computers), that teachers can implement within an hour in their classrooms, thus giving students a first taste of computing. Grown into a global movement, it has now reached over 1.8 billion activities completed by learners in 180+ countries and territories. In 2023, HoC explicitly invited activity providers to contribute AI/ML learning activities, offering an expanded set of AI activities for the annual Computer Science Education Week. In this paper, we take a closer look at the HoC activities we identified as dealing with AI/ML by addressing the following research questions: (1) How do HoC AI-related activities engage with the five big ideas? And (2) How do HoC AI-related activities engage with ideas related to ML? We conducted a content analysis of all 50+ HoC activities dealing with AI and ML. In the discussion, we provide recommendations for future development of introductory AI/ML activities.

\section{Background}

Efforts to broaden participation in computing—and by extension, in AI/ML education—have been led by government, university, and industry initiatives with the larger goal to make computing part of K–12 education for all \cite{vee2017coding}. As part of these efforts, the Silicon Valley-funded non-profit Code.org launched HoC in December 2013 during Computer Science Education Week. Over the years, HoC has grown to include hundreds of activities for different platforms, sometimes bundled with commercial content, and received widespread support from public figures such as Nobel Prize winner Malala Yousafzai, President Obama, and Pope Francis. As perhaps one of the most successful educational outreach campaigns, surprisingly little is known about its outcomes other than the staggering number of participants it has reached \cite{yauney2023systematic}.

More recently, several initiatives have shifted their attention to AI/ML applications, such as AI4ALL \cite{judd2020activities} and the Day of AI \cite{breazeal2023day}. Started in 2022 by MIT as an annual event, Day of AI has been implemented in over 7,500 classrooms and across 110 countries around the globe, supporting young people to better understand AI/ML and its impact \cite{hollands2024establishing}. These activities, arranged by specific grade ranges (e.g., elementary, middle, and high school), inform students about AI and possible careers in AI. Younger students can learn how machines learn from data using Teachable Machines \cite{carney2020teachable}, while older students get introduced to recommendation systems in social media, deep fakes, and image classification. Reports from participants revealed that these activities improved students’ understanding of AI \cite{hollands2024establishing}.

For this paper, we turn to HoC, which explicitly centered AI-related activities as part of 2023 Computer Science Education Week. Whereas prior to 2023 few activities included AI/ML content, the theme for the 2023 HoC was “Creativity with AI.” According to Hadi Partovi, CEO of Code.org, HoC activities are the beginning of longer computing and AI education in efforts “to prepare [students] for the future" by teaching "what an algorithm is, how the internet works, or how artificial intelligence is changing society” \cite{youtubeCode}. This paper builds on prior work that examined the conceptual, creative, and critical dimensions in the 340+ HoC activities offered in 2020 and 2021 \cite{morales-navarro2021investigating, morales-navarro2022iscomputational} to better understand what kind of content, perspectives, and engagement with AI and ML was promoted in the activities. Additionally, a prior study on HoC 2021 revealed that most activities (90\%) were designed for conceptual engagement with computing, but only 11\% provided opportunities for creative engagement. Less than 2\% promoted ethical or critical engagement with the discipline \cite{morales-navarro2022iscomputational}. Thus, designing introductory learning activities requires not only making important decisions on the tools and methods to scaffold learning about complex concepts and processes but also deciding how to integrate ethical and critical considerations into technical and functional aspects of computing, as well as making learning relevant to students’ interests and everyday lives.

\begin{figure}[t]
\centering
\includegraphics[width=0.9\columnwidth]{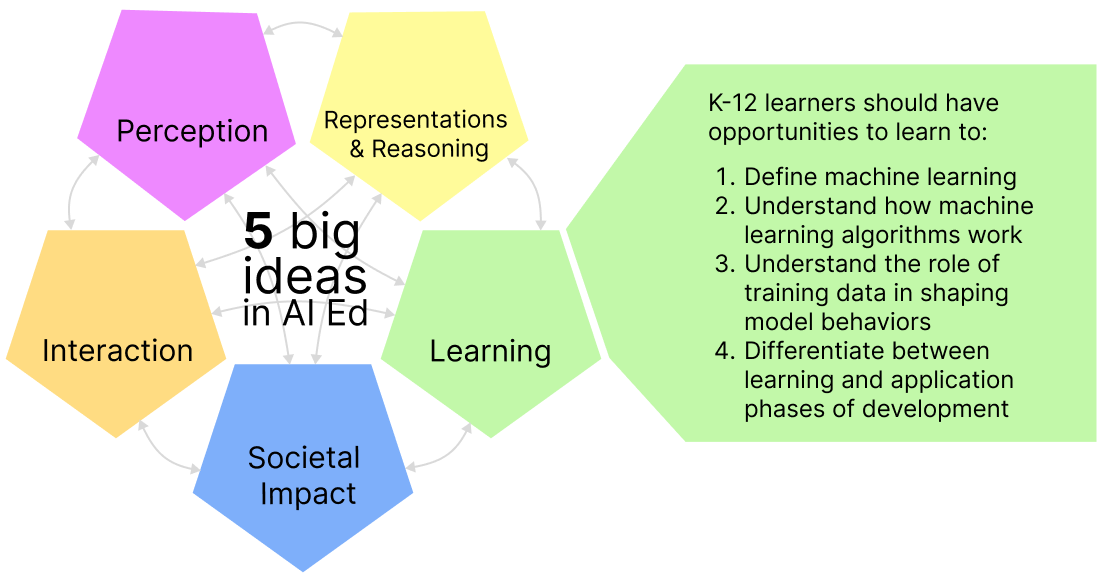} 
\caption{Diagram of the 5 big ideas in AI education proposed by Touretzky et al. \shortcite{touretzky2019envisioning} and (in green box) details about what ML learning activities should promote \cite{touretzky2023machine}}
\label{five}
\end{figure}

To gain a better understanding of how HoC activities engage with AI/ML, we used the five big ideas framework developed by Touretzky and colleagues \shortcite{touretzky2019envisioning} and examined how the five big ideas—(1) perception, (2) learning, (3) natural interaction, (4) representation and reasoning, and (5) societal impact—were integrated (see Figure \ref{five}. We investigated two of these ideas—learning and impact—in more detail. For ML, we turned to Touretzky and colleagues \shortcite{touretzky2023machine} discussion of ML which outlines the following five aspects: (1) defining ML; (2) understanding how ML algorithms work; (3) understanding the role of training data in shaping model behaviors; (4) understanding how data may be biased; and (5) knowing how to differentiate between learning and application phases of development. For societal and environmental impact, we investigated the different topics or issues addressed in learning activities. Finally, we explored the types of instructional approaches chosen to communicate these ideas and concepts to learners. The approaches ranged from telling students about ideas to involving them in building models, experimenting, and creating applications. In the latter, learners learn new concepts from labeled data, construct decision trees with labeled data, simulate how a neural network learns by adjusting its weights, explore historical datasets, and train models based on real-world datasets. 

\section{Methods}
\subsection{Context}
The context of this study is the Hour of Code (HoC), a repository created by Code.org hosting hundreds of different activities on hourofcode.com. These activities were primarily contributed by Code.org partners, including technology companies and educational organizations. The activities are categorized by grade ranges, beginner or advanced level, and by topic (e.g., art, media, music; artificial intelligence; language arts; science; social studies; computer science only). Many, but not all, activities have short teaser videos introducing the activities and provide instructional guides for how teachers can use them in their classrooms. Each year, in preparation for Computer Science Education Week in December, new activities are added. For instance, in 2020, the repository included 348 beginner activities for middle and high school youth \cite{morales2021investigating}, while in 2021, 495 activities were available for the same age ranges \cite{morales2022computational}.

\subsection{Data Collection}
In this study, we focused on publicly available HoC beginner activities dealing with AI/ML for middle and high school youth. Using the filtering feature of the HoC website (hourofcode.com), we selected beginner activities offered to middle and high school (grades 6–8 and grades 9+) students and teachers. In December 2023, 542 beginner activities were listed for middle and high school youth and by April 2024, the number had increased to 557 activities. We reviewed the short description of each activity to select those mentioning AI/ML and included all activities labeled as AI-related by the website in our analysis. All activities with working hyperlinks to activity details were included in our analysis, and those with broken links were excluded. This process resulted in 47 AI-related HoC activities for further analysis.

\begin{figure}[t]
\centering
\includegraphics[width=\columnwidth]{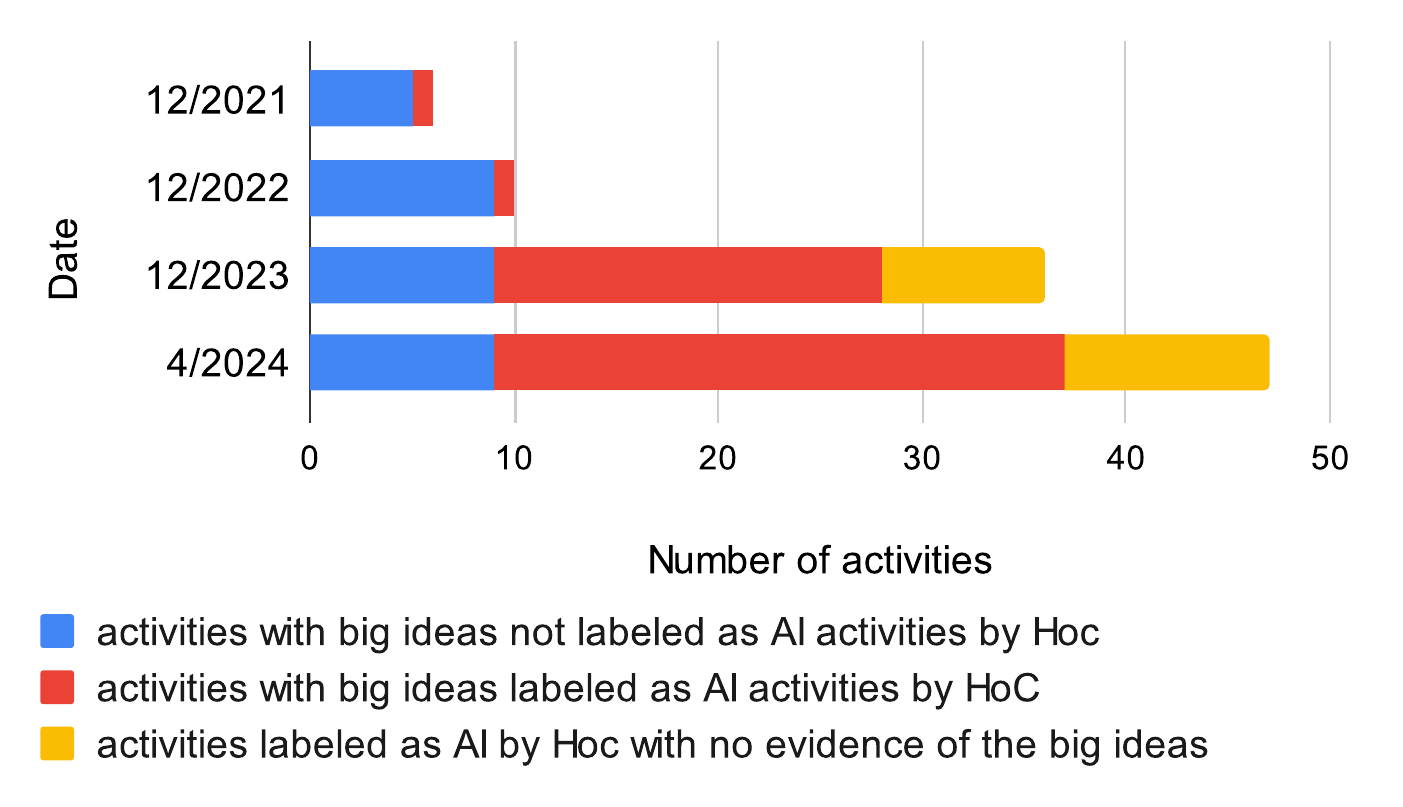} 
\caption{Number of AI-related activities in HoC by date. Color indicates if these activities were labeled as AI activities by HoC and if we found evidence of the five big ideas.}
\label{numbers}
\end{figure} 

\subsection{Analysis}
We analyzed the content of the selected 47 learning activities. These activities were coded deductively \cite{saldana2021coding} using the five big ideas about AI \cite{touretzky2019envisioning} as content categories, and the five aspects of ML \cite{touretzky2023machine} discussed above. Further, we inductively coded the societal impact topics covered in the activities and categorized the instructional methods showcased in the activity (e.g., hands-on, telling). Since the dataset was quite small, as it focused only on AI/ML-related activities, three researchers coded the activities collaboratively. 

First, two researchers completed each of the activities, taking notes on how these engaged with the five big ideas. Following, all researchers met, applied the coding scheme, and noted the different societal impact topics present in the activities. While coding, the researchers engaged dialogically with the data, seeking consensus and iteratively resolving disagreements with a fourth researcher who was familiar with the analysis. Because this is an exploratory study with a small number of learning activities, we prioritized establishing unanimous agreement on all coding (by coding collectively and resolving our divergent opinions through extensive discussion) over reliability (when coders apply the same scheme independently on the same data) \cite{mcdonald2019reliability}.

\section{Findings}
In the following sections, we provide an overview of HoC activities related to AI/ML, how activities engaged with the five big ideas of AI, and more specifically, how ML and societal impact were addressed.

\subsection{How Many HoC Activities Address AI and ML?}
As a first step, we examined HoC activity pools from 2021, 2022, 2023 (December), and 2024 (April) and established that the number of AI-related activities has increased rapidly from only six activities in 2021 to 10 in 2022, 36 in 2023, and 47 in 2024 (see Figure \ref{numbers}).

By April 2024, HoC offered over 557 beginner activities for middle and high school-aged learners. Only 38 (6.82\% of the total) of these activities were labeled as AI-related activities by HoC. Of these, 28 addressed at least one of the five big ideas in AI, while 10 did not involve any big ideas. For instance, “Pickcode: Generative Art Creator” guides learners to create art using loops and a function that returns random numbers. While the tutorial may be helpful for students to learn about variables, control flow statements, and functions, there is no AI/ML involved, and it does not engage with how generative models work and are trained. Another example is “Scratch - Book to Code,” where participants are guided to create a Scratch project about their favorite book by using the translation blocks available on Scratch. In this activity, there is no discussion of machine translation or AI/ML. Labeling activities as AI/ML that do not engage with the five big ideas as AI activities may be misleading for learners, educators, and parents. Further, through the analysis of activities not labeled as AI-related by HoC, we found nine other AI-related activities that engaged with the five big ideas.

\begin{figure}[t]
\centering
\includegraphics[width=\columnwidth]{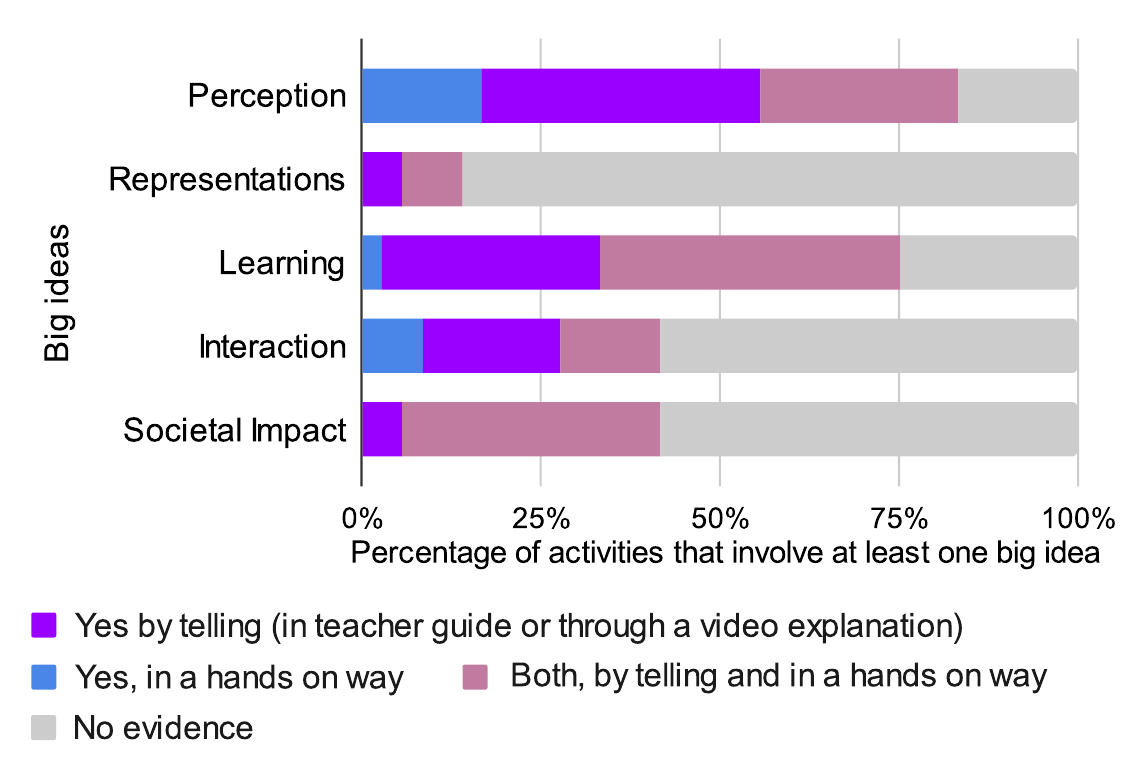} 
\caption{How HoC activities incorporate the big ideas. Colors indicate if big ideas were incorporated into hands-on activities, in telling activities, or both.}
\label{bigideas}
\end{figure}

\subsection{How Did AI HoC Activities Address the Five Big Ideas of AI?}

We then analyzed how these 37 activities (28 labeled by HoC, 9 through our own analysis of activity summaries) included the five big ideas (see Figure \ref{bigideas}). Of the activities that involved at least one big idea, perception was most commonly discussed (83.33\%), followed by learning (75\%), natural interaction (41.67\%), societal impact (41.67\%), and representation and reasoning (13.89\%). Thirty activities addressed perception, for example by including videos and slides that discuss how AI/ML systems make meaning of data collected using sensors in computer vision systems used in self-driving cars (e.g., AI with RVR+: Autonomous Vehicles), facial recognition (e.g., Old MacDonald Hacked a Farm, AI, AI Drone), and to recognize handwriting (e.g., Discover AI in Daily Life). 

Natural interaction was addressed in 15 activities that explicitly discussed human-computer interaction. Some of these activities presented examples of how humans can benefit from AI/ML systems, such as how image recognition applications can provide real-time descriptions to people who are blind (e.g., AI for Oceans). Other activities explained how people interact with everyday AI/ML-powered technologies, such as using facial recognition to unlock a phone or interacting with a voice assistant (e.g., Alexa in Space). Only five HoC activities integrated representation and reasoning. One of these activities, “Brain-in-a-bag”, was unplugged, involving youth in simulating artificial neural networks and comparing them to biological neural networks. Another activity, “How Convolutional Neural Networks Work,"  used video explanations to discuss representations and reasoning in the context of a computer vision system that recognizes handwritten digits. We discuss in more detail how activities integrate learning and societal/environmental impact below. 

Here, it is worth noting that ideas were often integrated into activities through videos or teacher explanations (as proposed in teacher guides) without opportunities for students to engage in hands-on ways. This was particularly the case for perception and learning. For instance, “AVATAR: Big Data \& Digital Footprints” addressed perception through a clickable tutorial where youth could click on different everyday technologies to learn about how they use sensors to collect data. “A.I. Rescue,” which simulates an artificial neural network included video explanations that addressed how neural networks learn from training data. Only several activities had learners design their own AI/ML applications. These applications often involved using off-the-shelf pre-trained models such as in “Weather Control” where youth could use pose recognition blocks to control the weather within a simulation. Other activities, such as “AI \& Neural Networks” involved youth in AI model training through Teachable Machines.

\begin{figure}[t]
\centering
\includegraphics[width=0.9\columnwidth]{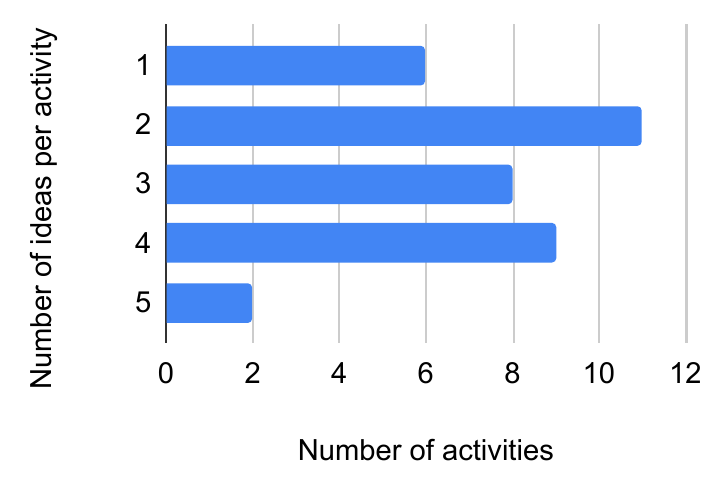} 
\caption{Distribution of activities by number of big ideas present in each activity.}
\label{ideas}
\end{figure}

Notably, only two activities, “Data Literacy in the AI Era Workshop” and “Build Your Own Chatbot in Python,” integrated all five big ideas (see Figures \ref{ideas}). The chatbot activity, for example, includes explanations about perception in relation to voice recognition. It also engages with  representations and reasoning through the discussion of rule-based systems, and includes explanations and activities about learning where youth train their chatbots with their own custom data. Furthermore,  it includes considerations of human interaction in the design of chatbots, and discussion of privacy concerns. While these HoC activities addressed all five ideas, they required a level of technical expertise and more than one hour.

\subsection{How Did AI HoC Activities Introduce ML?}

\begin{figure}[t]
\centering
\includegraphics[width=\columnwidth]{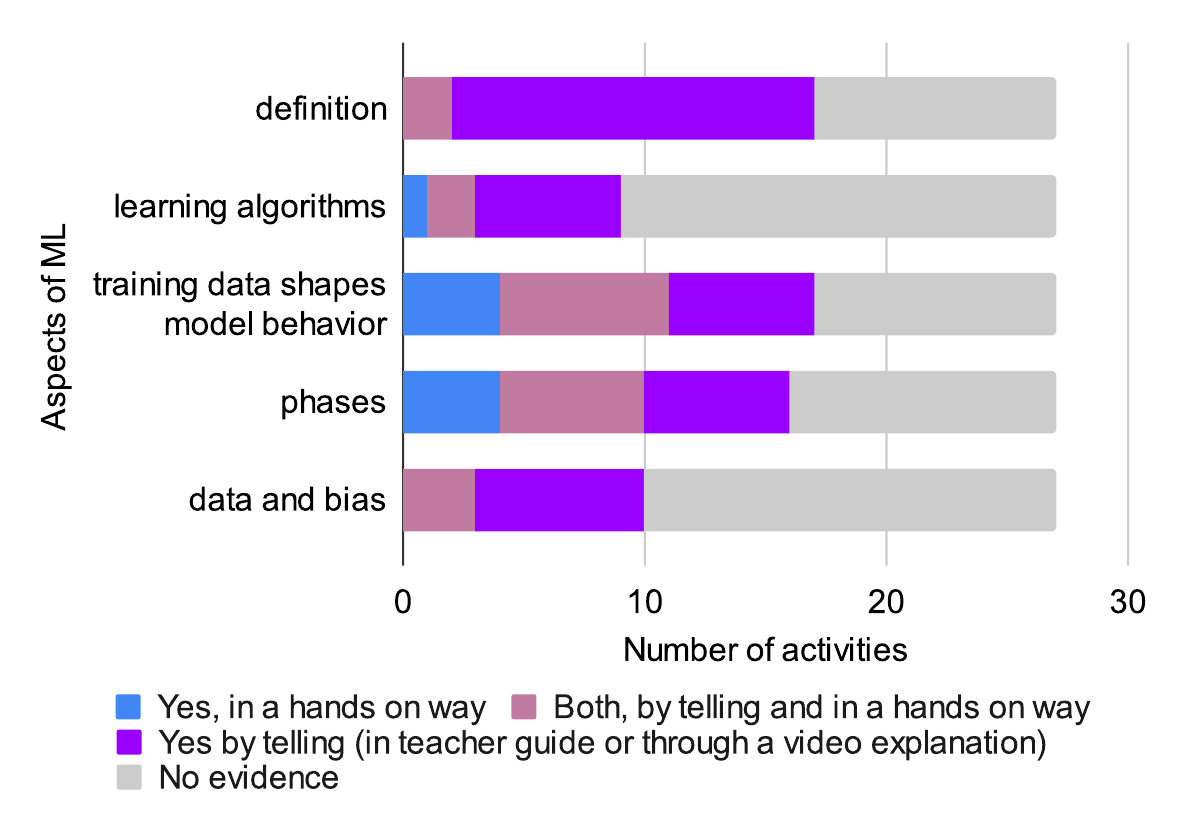} 
\caption{How HoC activities addressed different aspects of ML. Colors indicate if these aspects were incorporated into hands-on activities, in telling activities, or both.}
\label{ML}
\end{figure}

As noted earlier, 75\% of the activities that incorporated the big ideas addressed learning. These 27 activities introduced ML in diverse ways and centered on different aspects of ML (see Figure \ref{ML}). It is worth noting that ten of these 27 activities did not provide learners with a definition of ML. “Alexa in Space,” for example, addressed how voice assistants use ML models to recognize words, but did not define ML. Frequently, activities (17 activities) ascribed model behavior to data, introducing how training data shapes the way models work. “BOLT meets ChatGPT” included a video that explained that the behavior of transformer models is shaped by the data used to train them. "Driving the inteli AI Sign Train,” “Developing AI Literacy (DAILy) 2.0: Intro to Supervised Machine Learning,” and “AI for Oceans” involved learners in hands-on activities to create and modify training data sets to change model behaviors. “N-Gram Explorers: Quest for Words” is a tutorial that allowed learners to test different predetermined datasets to observe how the outputs changed. 

\begin{figure}[t]
\centering
\includegraphics[width=0.6\columnwidth]{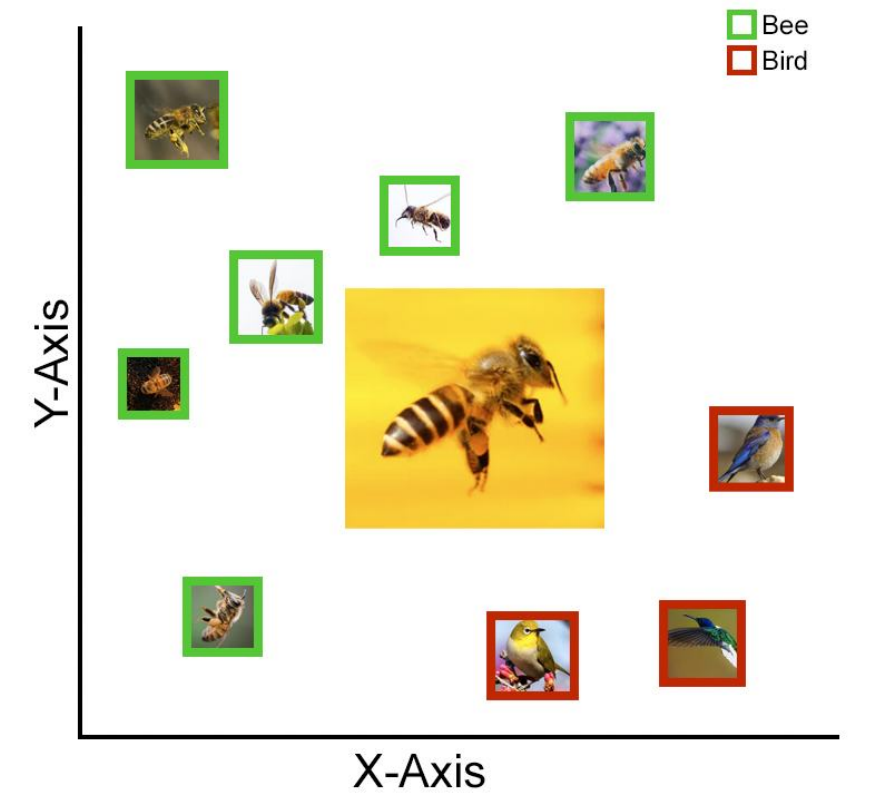} 
\caption{Machine Vision in Robotics (Virtual) uses bees and birds to explain K-nearest neighbors using a two-axis graph}
\label{bees}
\end{figure}

\begin{figure}[t]
\centering
\includegraphics[width=0.9\columnwidth]{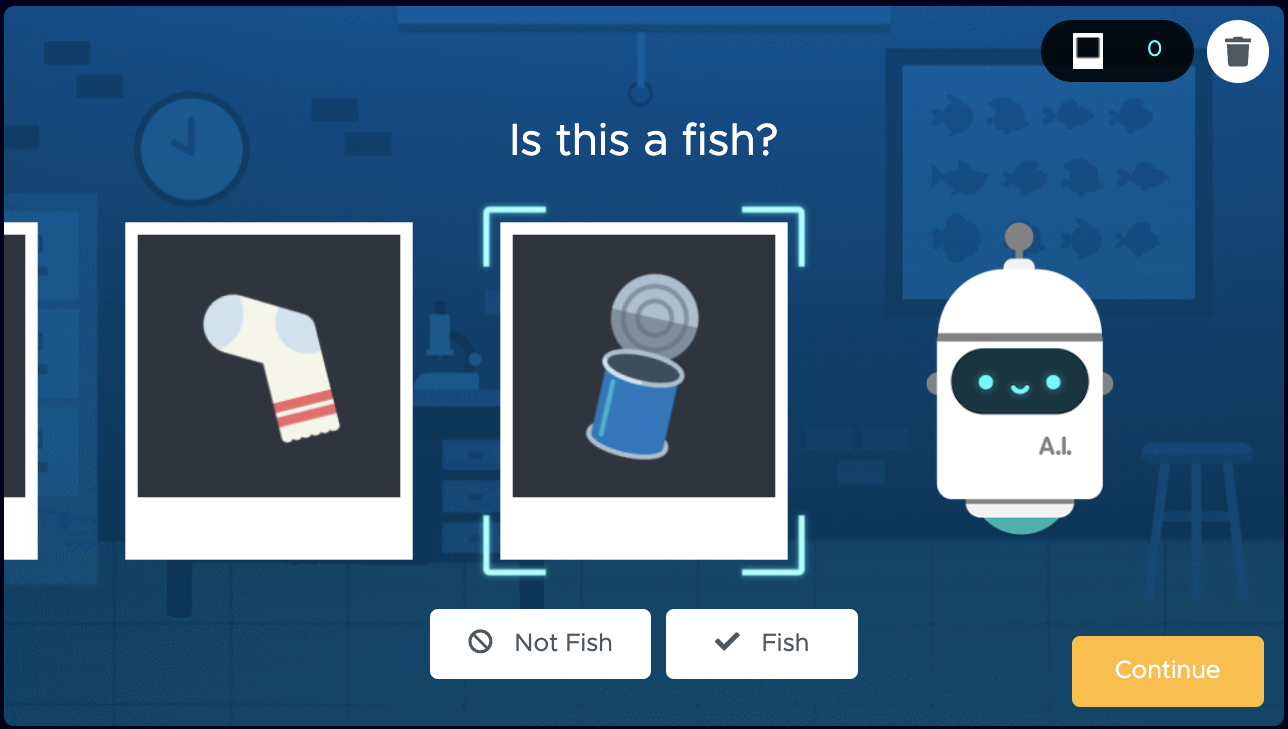} 
\caption{Data labeling interface in AI for Oceans}
\label{ocean}
\end{figure}

Less frequently, activities (9 activities) addressed learning algorithms and how these shape model behaviors. Six of these activities introduced learning algorithms through discussions. For example, “Machine Vision in Robotics (Virtual)” included a brief text explanation of K-nearest neighbors (see Figure \ref{bees}), while “AI4ALL: AI \& Dance” included a slideshow that explained that the models used in the activity were trained using neural networks and provided links to further details about how neural networks work. 16 activities introduced training, testing, and implementation phases of developing ML models. Activities such as “AI for Oceans” (see Figure \ref{ocean}) or “AI is a Hoot: Pose Detection” involved youth in training, and then testing classifiers. Ten activities addressed data and bias. “AI4ALL: AI \& Drawing” discussed how biases in training data can be reflected on outputs in the context of applications that recognize drawings, while “Face the Future” discussed how biases in facial recognition technologies, when used for policing, can perpetuate harm in already overpoliced communities. 

\subsection{How Did AI HoC Activities Address the Social/Environmental Impact of AI/ML?}

\begin{figure}[t]
\centering
\includegraphics[width=0.9\columnwidth]{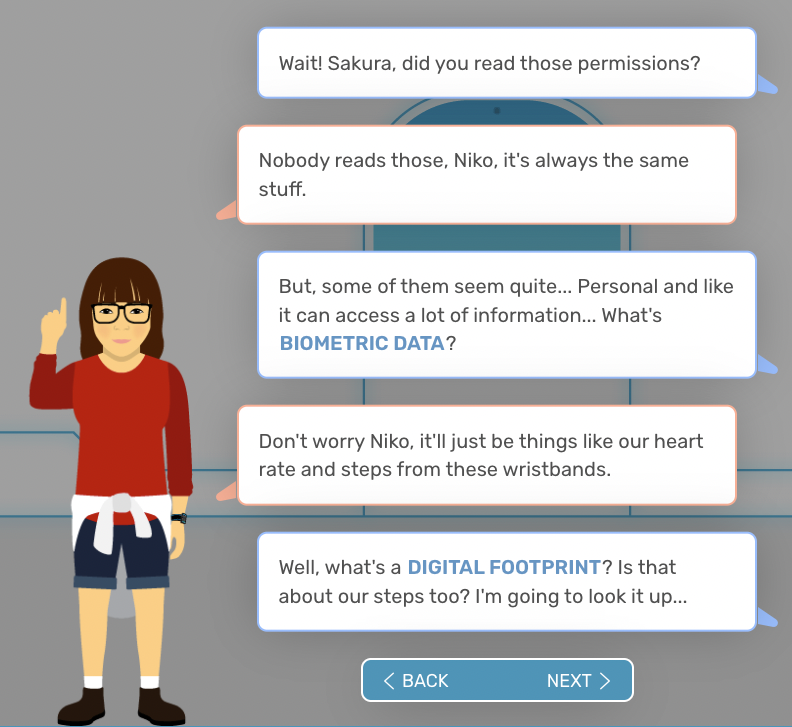} 
\caption{AVATAR: Big Data \& Digital Footprints discusses surveillance and the implications of sharing personal data.}
\label{avatar}
\end{figure}

Overall, fifteen activities (41.67\%) addressed the societal/environmental impact of AI/ML, introducing issues related to privacy, harmful biases, surveillance, misinformation, environmental impact, and ethics of self-driving cars. Here, eight of the activities focused on privacy issues. “Alexa in Space” addressed privacy concerns of using voice assistants, while “AVATAR: Big Data \& Digital Footprints” (see Figure \ref{avatar}) discussed how personal data is collected across applications and could be used to train models. Another eight activities discussed potential harmful biases. “Discover AI in Daily Life” discussed how resume screening systems may have harmful biases that prevent historically excluded communities from having access to jobs and “Generation AI” explored how harmful biases may affect people with disabilities. Furthermore, three activities discussed issues of surveillance. These include previously discussed activities “Alexa in Space,” “AVATAR: Big Data \& Digital Footprints,” and “Face the Future.” Two activities addressed misinformation. “Developing AI Literacy (DAILy) 2.0: Deepfakes + Ethical Matrix” and “BOLT meets ChatGPT” both addressed how synthetic text and synthetic images could be used to manipulate people and produce misinformation. One activity, “AI with RVR+: Automous Vehicles," discussed the ethics of self-driving cars using trolley problems. Finally, there were two activities that addressed environmental impact from a technosolutionist perspective. Here, “Old MacDonald Hacked a Farm, AI, AI Drone” addressed how AI/ML could reduce the environmental impact of farming through the development of more efficient techniques and less harmful fertilizers, while “AI for Oceans” highlighted how AI/ML systems could be used to address plastic pollution in the ocean.  

\section{Discussion}

In this paper, we examined the public portfolio of AI/ML activities for middle and high school students available in HoC. Upfront, we want to acknowledge the meteoric rise of AI and ML activities in the past three years, starting with only six activities in 2021 and rising to feature over 50 activities by April 2024. This is undoubtedly a significant increase, indicative of the importance AI/ML is now playing not only in computing education but in society at large. The HoC and its associated powerful public outreach play an important role given the urgent calls to support AI literacy for all. In the following discussion, we address in more detail what participants could take away from these offerings and how we can move forward with better designs. We do so with the understanding that further research should move beyond a content analysis to actually observing and evaluating how HoC activities are implemented with students. 

\subsection{Offering AI/ML Introductory Activities}

Our first discussion point focuses on the ‘What’—the content offerings. Using the Touretzky and colleagues’ \shortcite{touretzky2023machine} five big ideas, we found that perception was most commonly discussed, followed by learning. Other ideas such as natural interaction and societal impact were much less prominent, and representation and reasoning had the least attention. This finding is not surprising, given that some ideas are more accessible than others, whereas ideas such as representation and reasoning are more difficult due to required background knowledge. We also noted an uneven level of quality in the offered HoC activities. Furthermore, some activities—as noted above—were labeled as AI when they clearly did not engage with any of the five big ideas. Finally, several activities were not accessible to novice learners, requiring substantial programming background. While some of these observations might be due to the novel nature of AI/ML in K–12 education, this presents misleading information to learners, educators, and parents. We have to build out a knowledge base and instructional approaches to make these concepts accessible. 

A second discussion point concerns the ‘Why’—the societal impact. We noticed increased attention paid to societal impact. Nearly one-third of all AI/ML-related HoC activities addressed some form of societal impact. This is a great improvement, considering that earlier research on HoC activities showed that less than 2\% of activities in previous years engaged with critical issues in computing \cite{morales-navarro2021investigating, morales-navarro2022iscomputational}. This seems to reflect a growing trend to address critical issues with AI, more so than with computing in general. In this context, we also noticed a problematic position in what concerns environmental impact. Two HoC activities approached AI/ML from a technosolutionist perspective\cite{costanza2020design}, ignoring the increased challenges in energy consumption used to train AI/ML models \cite{luccioni2023estimating}. 

A third discussion point concerns the ‘How’—the instructional approaches. We noted how much more “telling” than “doing” (or hands-on inquiry) was chosen in HoC activities. This is a surprising finding considering the compelling evidence from decades of computing education research on the important role designing applications plays in supporting novices \cite{oleson2021onthe, waite2021teaching, kafai1991learning}. One possible explanation is rooted in the nature of HoC activities, which have been known to focus less on creative engagement \cite{morales-navarro2021investigating} favoring a more puzzle-like approach in their designs, as some other critics have noted \cite{yauney2023systematic}. While the time period of one hour certainly limits investigations and designs more commonly associated with project-based activities, there are some compelling hands-on tools such as Teachable Machines \cite{carney2020teachable} which have succeeded in making some aspects of ML salient to even young learners \cite{williams2019aisfor}. Another possible explanation is the fairly limited research base on how to teach AI and ML in K–12 education. As noted above, current national frameworks and standards do not explicitly include these topics, and we need to build a better understanding of how AI/ML can be taught in such a fashion and develop a more extensive repository of tools to support learners in building applications.

\subsection{Designing Introductory Learning Activities for AI and ML}

Notwithstanding our concerns about the HoC activities, we do think that introductory activities have a place in the current educational landscape, providing learners with a variety of entry points into AI and ML. Based on previous analyses of HoC’s potential for conceptual, creative, and critical engagement \cite{morales2022computational}, we offer several recommendations that can expand the scope, modalities, focus, and tools with AI/ML of future HoC activities.

\subsubsection{Addressing the Full Spectrum of AI/ML Ideas}

As a first step, it is necessary to broaden the range of AI ideas addressed in HoC. Furthermore, it is important to be clear in what is AI and what it is not, and that generative AI is a subarea but not all what AI is. Here it would also be important to provide basic and clear definitions. HoC activities could also tackle perhaps less accessible but nonetheless important ideas and thus help broaden youth’s understanding of AI/ML. 

\subsubsection{Offering Unplugged and Collaborative AI/ML Activities}

There are already numerous examples of unplugged activities for understanding various AI concepts such as facial recognition \cite{lim2024unplugged}, decision trees and reinforcement learning \cite{lindner2019unplugged}, ethics \cite{ali2019constructionism}, or semantic networks and knowledge representations \cite{long2021co}. Many of these do not require any prior understanding and are accessible to learners of all ages and thus could be repurposed for HoC activities. Furthermore, all the HoC activities were individual in nature and, while presumably implemented in a classroom context, neglected to take into account the social nature of data production and curation and design of computational systems \cite{tseng2024co, hardy2020data}.

\subsubsection{Going Beyond the Data-Centric Focus of Current AI/ML Activities}

Like in many of the currently designed AI and ML learning activities, we noticed an overreliance on data-centric approaches, with little to no attention paid to the learning algorithms used in AI/ML systems \cite{morales2024unpacking}. We suggest that designers of learning activities also direct their attention to this neglected part of learning about AI/ML and develop approaches on how to help K–12 students learn about learning algorithms. While many of these are challenging topics, some requiring understanding of advanced mathematical concepts, it is not impossible, as Broll \& Grover \shortcite{broll2023beyond} have shown in their research studies.

\subsubsection{Integrating Societal Impact Into AI/ML}

The current set of HoC activities focused on AI and ML already does pay more attention to societal impact than previous HoC activities focused on computing at large \cite{morales2022computational}. Rather than treating societal impact as a stand-alone topic, here the integration with big ideas could be improved so students learn about key concepts and challenges at the same time. Furthermore, topics that address ethics and bias often focus on high-stakes issues such as self-driving car trolley problems while neglecting to address the everyday occurrences of potentially harmful algorithmic bias in web searches, and social media applications. While both types of issues are important, for many youth, the latter cases are the ones they might encounter in their everyday interactions with technology.

\subsubsection{Developing More Accessible Tools for Hands-On AI/ML}

Tools designed for novices to build models and explore how models work by visualizing their inner workings could make learning about AI/ML much more feasible for novice learners \cite{gresse2021visual}. Beyond the successful Teachable Machine \cite{carney2020teachable} we need better tools that can support learners to not just use AI/ML but create AI/ML models—an approach that would be very much aligned with the theme of Code.org’s HoC “Creativity with AI.”

\section{Conclusions}
In this study, we reported on HoC activities offered in 2023 to introduce middle and high school youth to  artificial intelligence and machine learning. The content analysis revealed that the 47 HoC activities concentrated on a subset of big ideas in AI and ML with greater attention paid to critical and ethical issues. While only a small number of activities provided hands-on experiences, other activities required advanced programming knowledge to complete. In addition, many activities labeled as AI/ML in fact included little to no information on the topic. In our discussion, we provide several recommendations on how introductory activities like the HoC can be improved in their design and scope to leverage their broad outreach.

\section{Acknowledgments}
With regards to Daniel Noh for his feedback.

\bibliography{aaai25}

\end{document}